\documentclass[aps,twocolumn,showpacs]{revtex4}

\usepackage{amssymb}

   \newcommand*{\bx}{\mathbf{x}}
   \newcommand*{\al}{\alpha}
   \newcommand*{\da}{\dagger}
   \newcommand*{\ep}{\epsilon}
   \newcommand*{\et}{\eta}
   
   \newcommand*{\nn}{\nonumber}
   \newcommand*{\ps}{\psi} 
   \newcommand*{\si}{\sigma} 
   \newcommand*{\De}{\Delta}                                          
   \newcommand*{\Eq}[1]{Eq.~(\ref{eq:#1})}
   \newcommand*{\eq}[1]{(\ref{eq:#1})}
   \newcommand*{\bracket}[1]{\langle#1\rangle}

\begin{document}
\title{Physical content of Heisenberg's uncertainty relation: 
Limitation and reformulation}

\author{Masanao Ozawa}

\affiliation{Graduate School of Information Sciences,
T\^{o}hoku University, Aoba-ku, Sendai,  980-8579, Japan}

\begin{abstract}
Heisenberg's reciprocal relation between position measurement error and
momentum disturbance is rigorously proven under the assumption  that
those error and disturbance are independent of the state of the measured
object. A generalization of Heisenberg's relation proven valid  for
arbitrary measurements is proposed and reveals two distinct types of
possible violations of Heisenberg's relation.
\end{abstract}
\maketitle

\section{Introduction}

An essential departure of quantum mechanics from classical
mechanics is that any measurement of a microscopic object
involves the interaction with the apparatus not to be neglected
and accordingly introduces an unavoidable and uncontrollable
disturbance on the measured object.
Undoubtedly, this point of view led to fundamental doctrines of 
the Copenhagen interpretation of quantum mechanics
\cite{Boh28}, which successfully dissolved the wave-particle 
duality and the continuous-discontinuous discrepancy.
In his celebrated paper \cite{Hei27} published in 1927, 
Heisenberg attempted 
to establish the quantitative expression of the amount of
unavoidable momentum disturbance caused by any position 
measurement.
His statement, with some elaborations,  can be formulated
as follows:
{\em For every measurement of the position $Q$ of a mass with 
root-mean-square error $\ep(Q)$, the root-mean-square
disturbance $\et(P)$ of the momentum $P$ of the mass caused 
by the interaction of this measurement always satisfies the 
relation}
\begin{equation}\label{eq:Heisenberg}
\ep(Q)\et(P)\ge \frac{\hbar}{2}.
\end{equation}

Heisenberg \cite{Hei27} not only explained the physical intuition 
underlying the above relation by discussing the $\gamma$ 
ray microscope thought experiment, but also claimed that this 
relation can be proven as a straightforward consequence of the 
canonical commutation relation (CCR), $QP-PQ=i\hbar$.  
A mathematical part of his proof was refined by introducing
the notion of standard deviation shortly afterward by Kennard 
\cite{Ken27} and later generalized to arbitrary pair of observables
by Robertson \cite{Rob29} as the following statement:
{\em For any pair of observables $A$ and $B$, their standard 
deviations, $\si(A)$ and $\si(B)$, satisfy the relation
\begin{equation}\label{eq:Robertson}
\si(A)\si(B)\ge \frac{1}{2}|\bracket{\ps,[A,B]\ps}|
\end{equation}
in any state $\ps$ with $\si(A), \si(B)<\infty$.}
In the above, $[A,B]$ stands for the commutator
$[A,B]=AB-BA$, and the standard deviation is defined as
$\si(A)=(\bracket{\ps,A^{2}\ps}-\bracket{\ps,A\ps}^{2})^{1/2}$,
where $\bracket{\cdots,\cdots}$ denotes the inner product \cite{note-1}.
As a consequence of $[Q,P]=i\hbar$, we have
\begin{equation}\label{eq:Kennard}
\si(Q)\si(P)\ge \frac{\hbar}{2},
\end{equation}
which was proven by Heisenberg himself for Gaussian states
and by Kennard \cite{Ken27} generally.

Heisenberg \cite{Hei27} argued that the mathematical relation 
\Eq{Kennard} concludes the physical assertion expressed by
\Eq{Heisenberg}.   Since then, his claim has been accepted by many
\cite{vN32,Boh49,Boh51,Mes59a,Bra74,CTDSZ80,BVT80}.
However, the universal validity of \Eq{Heisenberg} has been also
criticized in many ways 
\cite{EPR35,AK65,Bal70,Yue83,Kra87,AG88,88MS,89RS,%
HV90,MM90,91QU,Ish91,BK92,MM92,App98,01CQSR,02KB5E}.
In fact, the ``resolution power'' of the $\gamma$ ray microscope cannot be
identified in any interpretation with the standard deviation of the mass 
position in the state to be measured.  Thus, \Eq{Kennard} cannot be 
considered as a ``formal expression'' of \Eq{Heisenberg}.  Moreover, 
no one has succeeded in proving \Eq{Heisenberg} for general measurements
even by applying \Eq{Kennard} not only to the mass state but also to the
apparatus state. Undoubtedly, this has caused serious confusions among
physicists on the status of this leading principle of quantum mechanics.   

In order to clarify the confusion, we shall start with precise definitions
of root-mean-square disturbance and mean-square error in arbitrary
measurements and reconstruct Heisenberg's argument. His argument
applies \Eq{Kennard} to the mass state just after the measurement.
However, in order to derive \Eq{Heisenberg}, additional assumptions
are necessary. Here, we shall show that the following two assumptions
completes Heisenberg's argument: (i) Both amounts of $\ep(Q)$ and
$\et(P)$ are independent of the input state.  (ii) The measurement
always leaves the mass with position standard deviation smaller  than
the $\ep(Q)$. Thus, Heisenberg's argument left open the following
questions: Can we further relax the assumptions? What relation holds
for arbitrary measurements?  What conditions characterize violations
of \Eq{Heisenberg}? In particular, the second assumption, which will
be called the equipredictivity of measurement, stringently restricts the
class of measurements to which 
\Eq{Heisenberg} is applicable. In order to answer those questions, a
new approach to proving \Eq{Heisenberg} will be proposed based on
analysis of commutation relations for the operators representing the
error and the disturbance. From this approach, we shall obtain a
generalization of  Heisenberg's relation proven valid for arbitrary
measurements and also obtained a new proof of Heisenberg's relation
without assuming the equipredictivity. We shall further discuss
limitations on Heisenberg's relation  based on the generalized relation
and show that there are two distinct types  of measurements in which
Heisenberg's relation is violated uniformly for any input state. The
significance of the results will be discussed in the final section. In this
Letter, we shall confine our attention to the position measurement
error and momentum disturbance; generalizations to arbitrary
observables are partly discussed in a separate paper
\cite{03UVRHUP}. 

\section{Formulation of error and disturbance}

Let us consider a measurement of position $Q$ of a mass
with momentum $P$.
The interaction between the mass and the apparatus is assumed 
to turn on at time 0 and turn off at time $\Delta t$.
Let $\ps$ be the input state, i.e., the state of the mass 
just before the interaction, 
and let $\xi$ be the state of the apparatus just before 
the interaction.
Let $U$ be the unitary operator representing the time
evolution of the mass plus apparatus for the time interval
$(0,\De t)$.
Then, in the Heisenberg picture the momentum change 
is represented by
\begin{equation}\label{eq:disturbance}
D(P)=P(\Delta t)-P(0),
\end{equation}
where $P(0)=P\otimes I$ and $P(\De t)=U^{\da}P(0)U$.
The root-mean-square momentum disturbance is defined by
\begin{equation}
\et(P)=\bracket{D(P)^{2}}^{1/2},
\end{equation}
where $\bracket{\cdots}$ denotes the mean value in the
original state $\ps\otimes\xi$.

Now we shall discuss some basic properties of the above
quantity.
The above statistical definition leads to the following
geometric expression 
\begin{equation}
\et(P)=\|P(\De t)\ps\otimes\xi-P(0)\ps\otimes\xi\|.
\end{equation}
Thus, by the relation
\begin{equation}
\si(A)=\|A\ps\otimes\xi-\bracket{A}\ps\otimes\xi\|
\end{equation}
for any observable $A$, 
we easily obtain the following relation
\begin{equation}\label{eq:bound}
|\si(P(\De t))-\si(P(0))|\le\et(P)+|\bracket{P(\De t)}-\bracket{P(0)}|,
\end{equation}
stating that the change in the momentum standard deviation is 
bounded from above by the root-mean-square disturbance except 
for the change in the mean momentum. 
If the mass had a definite momentum $p$ before the measurement,
i.e., $P(0)\ps=p\ps$, we would have 
\begin{equation}
D(P)\ps\otimes\xi=[P(\De t)-p]\ps\otimes\xi,
\end{equation}
and hence
\begin{equation}
\et(P)^{2}=\bracket{[P(\De t)-p]^{2}}\ge\si(P(\De t))^{2}.
\end{equation}
Thus, the momentum standard deviation after the measurement is
bounded from above by the root-mean-square disturbance.
In particular, if $p=0$, we have
\begin{equation}\label{eq:P-disturbance}
\et(P)^{2}=\bracket{P(\De t)^{2}}.
\end{equation}
This relation can be interpreted as follows. {\em If the mass is at rest
before the  measurement, all the object's momentum after the 
measurement arises
from the interaction, so that the mean-square disturbance 
(i.e., the square of the 
root-mean-square disturbance) is equal to the mean-square momentum 
after the measurement.}

The role of the interaction is to transduce the value of $Q$ 
before the interaction to the value of an observable $M$ 
of the probe, a part of the apparatus, after the interaction.
We shall call $M$ the meter observable.
We suppose that after the interaction is turned off, 
the outcome of the $Q$ measurement in the state $\ps$
is obtained by measuring $M$ 
without further disturbing the momentum $P$ of the mass;
this is possible by another measuring apparatus coupled only 
to the probe.
The postulates of quantum mechanics do not limit the
accuracy of the latter measurement of $M$, 
and hence we neglect the error from this measurement.
Then, in the Heisenberg picture the error caused by
this process of the $Q$ measurement is represented by
\begin{equation}\label{eq:error}
N(Q)=M(\De t)-Q(0),
\end{equation}
where $Q(0)=Q\otimes I$ and $M(\De t)=U^{\da}(I\otimes M)U$.
The root-mean-square error is defined by
\begin{equation}
\ep(Q)=\bracket{N(Q)^{2}}^{1/2}.
\end{equation}
The above statistical definition leads to the following
geometric expression 
\begin{equation}
\ep(Q)=\|M(\De t)\ps\otimes\xi-Q(0)\ps\otimes\xi\|.
\end{equation}
Analogously with \Eq{bound}, we obtain
\begin{equation}\label{eq:bound}
|\si(M(\De t))-\si(Q(0))|\le\ep(Q)+|\bracket{M(\De t)}-\bracket{Q(0)}|,
\end{equation}
showing that the change in the standard deviation from input to output
is bounded from above by the root-mean-square noise except for
the bias, namely, the change in the mean value.
If the mass had a definite position $q$ before the 
measurement, we would have
\begin{equation}
\ep(Q)^{2}=\bracket{[M(\De t)-q]^{2}}\ge\si(M(\De t))^{2},
\end{equation}
so that $\ep(Q)$ would represent the root-mean-square deviation of 
the measurement outcome from the true position $q$ and
would limit the standard deviation of the outcome of the $Q$ 
measurement. 

\section{Reconstruction of Heisenberg's argument}

In the 1927 paper, Heisenberg claimed that \Eq{Heisenberg} is  a
straightforward mathematical consequence of the CCR, 
$QP-PQ=i\hbar$.   Although he gave a claimed proof of
\Eq{Heisenberg} from the CCR, his argument includes some implicit
assumptions.  Consequently, \Eq{Heisenberg} cannot be considered as
a straightforward  consequence of the CCR nor as a universally valid
relation. Heisenberg's idea of the proof is as follows:  Start with a
measurement with error
$\ep(Q)$, claim that the state just after the measurement has
$\si(Q)=\ep(Q)$, prove \Eq{Kennard} from the CCR for the above
state to obtain the relation
$\ep(Q)\si(P)\ge\hbar/2$, and identify $\si(P)$ with $\et(P)$ to obtain
\Eq{Heisenberg}. In order to clarify the hidden assumptions, in what
follows, we shall show that Heisenberg's original proof can be
rigorously reconstructed under the following two additional
assumptions:   (i) Both $\ep(Q)$ and $\et(P)$ are independent of the
input state; in this case, we say that the measurement has {\em constant
mean-square noise and disturbance}.   (ii) The measurement always
leaves the mass  with position standard deviation smaller than
$\ep(Q)$;  in this case, we say that the measurement is  {\em
equipredictive}.

Under the above assumptions, 
the proof of \Eq{Heisenberg} runs as follows.
In order to obtain an estimate of the momentum disturbance,
we shall consider the case where the mass were at rest before the
measurement.
Then, by \Eq{P-disturbance} the object's mean-square
momentum after the measurement
is equal to the mean-square momentum disturbance, i.e., 
\begin{equation}\label{eq:P-disturbance2}
\et(P)^{2}=\bracket{P(\De t)^{2}}.
\end{equation}
Let $\ps_{x}$ be the state of the mass after the 
measurement with outcome $x$.
We shall denote by $\si_{x}$ the standard deviation in the state $\ps_{x}$.
We shall later show that the relation
\begin{equation}\label{eq:P-SD}
\et(P)\ge\si_{x}(P)
\end{equation}
holds with positive probability.
On the other hand, by condition (ii), we have 
\begin{equation}\label{eq:Q-SD}
\ep(Q)\ge\si_{x}(Q)
\end{equation}
holds with probability one.
It follows from Eqs.~\eq{P-SD} and \eq{Q-SD}
that there must be a state vector $\ps_{x}$ satisfying
\begin{equation}
\ep(Q)\et(P)\ge\si_{x}(Q)\si_{x}(P).
\end{equation}
Thus, \Eq{Heisenberg} follows from \Eq{Kennard}, if the mass is 
at rest just before the measurement. Then, assumption (i) 
ensures that \Eq{Heisenberg} holds irrespective of the 
particular choice of the input state.   
We, therefore, conclude that {\em every equipredictive 
measurement with constant mean-square noise and
disturbance satisfies \Eq{Heisenberg} for every input state \cite{note-2}.}

Now, we shall prove \Eq{P-SD}.
The state $\ps_{x}$ arises with the probability distribution $\pi(dx)$ 
of obtaining the outcome $x$ determined by
\begin{equation}
\pi(dx)=\bracket{E^{M(\De t)}(dx)},
\end{equation}
where $E^{M(\De t)}$ is the 
resolution of the identity corresponding to the observable $M(\De t)$.
Then, since $\bracket{P(\De t)^{2}}$ is the mean of $P^{2}$ after the
measurement, we have
\begin{equation}\label{eq:P-MS1}
\bracket{P(\De t)^{2}}= \int\,\bracket{\ps_{x},P^{2}\ps_{x}}\,\pi(dx).
\end{equation}
Since $\bracket{\ps_{x},P^{2}\ps_{x}}\ge\si_{x}(P)^{2}$ for any $x$, 
we have
\begin{equation}
\int\,\bracket{\ps_{x},P^{2}\ps_{x}}\,\pi(dx)\ge
\int\,\si_{x}(P)^{2}\,\pi(dx),
\end{equation}
and consequently
\begin{equation}\label{eq:int-ineq}
\bracket{P(\De t)^{2}}\ge
\int\,\si_{x}(P)^{2}\,\pi(dx).
\end{equation}
Thus,  we have
\begin{equation}
\bracket{P(\De t)^{2}}\ge \si_{x}(P)^{2},
\end{equation}
with positive probability; otherwise the opposite inequality of
\Eq{int-ineq} would hold. This concludes the proof of  \Eq{P-SD}.

\section{Universally valid reformulation}

The above proof of \Eq{Heisenberg} rigorously reproduces
Heisenberg's intention in 1927 that the uncertainty relation expressed by
\Eq{Heisenberg} is a straightforward consequence from \Eq{Kennard} 
for the state after the measurement. 
However, the above proof does not show that \Eq{Heisenberg} is universally 
valid and has left the problem quite open as to the limitation of \Eq{Heisenberg}. 
Thus, the following problems have remained open on  Heisenberg's
uncertainty relation.
Can we further relax the assumptions of the proof?
What relation holds for arbitrary measurements?
What conditions characterize violations of \Eq{Heisenberg}?

In what follows,  we shall prove that 
{\em the relation 
\begin{equation}\label{eq:Ozawa}
\ep(Q)\et(P)+\ep(Q)\si(P)+\si(Q)\et(P)\ge\frac{\hbar}{2}
\end{equation}
holds for every measurement and every input state 
as long as all the relevant terms are finite},
where $\si(Q)$ and $\si(P)$ refer to the standard deviations
of $Q$ and $P$ in the input state.

The proof of \Eq{Ozawa} runs as follows.
Since $M$ and $P$ are observables in different systems, we have
$[M(\Delta t),P(\Delta t)]=0$.  
Substituting $M(\De t)=Q(0)+N(Q)$ and $P(\De t)=P(0)+D(P)$ 
for this relation and using the commutation relation 
$[Q,P]=i\hbar$,
we have \cite{AG88,91QU,Ish91}
\begin{equation}\label{eq:commutation}
[N(Q),D(P)]+[N(Q),P(0)]+[Q(0),D(P)]=-i\hbar.
\end{equation}
Taking the moduli of means of the both sides
and applying the triangular inequality, we have
\begin{eqnarray}\label{eq:in-1}
\lefteqn{
|\bracket{[N(Q),D(P)]}|+|\bracket{[N(Q),P(0)]}|}
\hspace{10em}\nn\\
& &+|\bracket{[Q(0),D(P)]}|\ge\hbar.
\end{eqnarray}
Since the variance is not greater than the mean-square,   
\Eq{Robertson} gives
\begin{equation}\label{eq:in-2}
\ep(Q)\et(P)\ge\si[N(Q)]\si[D(P)]
\ge\frac{1}{2}|\bracket{[N(Q),D(P)]}|,
\end{equation}
where $\si$  in the middle term refers to the standard deviation 
in the state $\ps\otimes\xi$.
Similarly, we have
\begin{eqnarray}
\ep(Q)\si(P)&\ge&\si[N(Q)]\si[P(0)]
\ge\frac{1}{2}|\bracket{[N(Q),D(P)]}|,\quad\quad\label{eq:in-3}\\
\si(Q)\et(P)&\ge&\si[Q(0)]\si[D(P)]
\ge\frac{1}{2}|\bracket{[Q(0),D(P)]}|.\quad\quad\label{eq:in-4}\
\end{eqnarray}
Thus, from Eqs.~\eq{in-1}--\eq{in-4}, we conclude \Eq{Ozawa}.

\section{Limitation of Heisenberg's relation}

The above proof clearly answers the question when 
\Eq{Heisenberg} holds generally: {\em If $[N(Q),P(0)]+[Q(0),D(P)]=0$,
then \Eq{Heisenberg} holds for every input state $\ps$
and every apparatus state $\xi$.}
In fact, from \Eq{commutation}, the condition  
$[N(Q),P(0)]+[Q(0),D(P)]=0$ implies the commutation relation
\begin{equation}\label{eq:commutation-1}
[N(Q),D(P)]=-i\hbar.
\end{equation}
Thus, \Eq{Heisenberg}  follows immediately from \Eq{in-2}. 

In order to draw a useful conclusion,
we say that the measurement has {\em independent intervention},
if there are two operators $N$ and $D$ on the state space of 
the apparatus such that 
\begin{eqnarray}
N(Q)&=&I\otimes N,\label{eq:ii-1a}\\
D(P)&=&I\otimes D,\label{eq:ii-1b}.
\end{eqnarray}
In this case, we have $\ep(Q)=\|N\xi\|$ and $\et(P)=\|D\xi\|$,
so that every measurement with independent intervention has
constant mean-square noise and disturbance.

From Eqs.~\eq{ii-1a} and \eq{ii-1b} the relations 
\begin{equation}\label{eq:ii-3}
[N(Q),P(0)]=[Q(0),D(P)]=0
\end{equation} 
follows easily.
Thus, we conclude that {\em every measurement 
with independent intervention 
satisfies \Eq{Heisenberg} for every input state $\ps$ 
and every apparatus state $\xi$
without requiring the equipredictivity.}  

The above conclusion suggests that the equipredictivity is not an essential 
requirement for a measurement to satisfy Heisenberg's relation
compared with the independence of the noise and the disturbance from
the measured system.

\section{Von Neumann models}

Von Neumann \cite{vN32} constructed the first quantum 
mechanical model of position measurement.  His model is 
described by the Hamiltonian 
\begin{equation}\label{eq:Neumann}
H=K(Q\otimes P_{0})
\end{equation}
that couples the mass position $Q$ and the probe momentum
$P_{0}$ with coupling constant $K$ satisfying $K\De t =1$.
Then, taking the meter to be the probe position, i.e., $M=Q_{0}$,
the time evolution $U=\exp(-i\De t H/ \hbar)$ determines the
input-output relation \cite{02KB5E}
\begin{eqnarray} 
M(\De t)&=&Q(0)+(I\otimes Q_{0}), \\ 
P(\De t)&=&P(0)-(I\otimes P_{0}).
\end{eqnarray} 
Thus, this model has independent intervention with
$N(Q)=I\otimes Q_{0}$ and $D(P)=-I\otimes P_{0}$,
and hence \Eq{Heisenberg} holds.
Assuming that the probe's mean position is initially at the origin, 
the constant root-mean-square error $\ep(Q)=\si(Q_{0})$ 
plays an analogous role of ``resolution power'' of microscopic
measurements.  
Then, \Eq{Heisenberg} follows directly from the
relation $\si(Q_{0})\si(P_{0})\ge\hbar/2$ \cite{02KB5E}.

\section{Equivalence of measurements with independent
intervention}

Interestingly, we can show that every measurement 
with independent intervention has the position error operator
and the momentum disturbance operator equivalent to 
those of the above von Neumann model. 
In fact, from Eqs.~\eq{commutation-1},\eq{ii-1a}, and \eq{ii-1b}, 
we have
\begin{equation}
[N,-D]=i\hbar.
\end{equation}
Thus, by the Stone-von Neumann theorem on the uniqueness
of representations of the canonical commutation relations
\cite{RS75}, 
the independent error and disturbance operators can be,
up to unitary equivalence, decomposed as 
\begin{eqnarray}
N&=&Q_{0}\otimes I,\\ 
-D&=&P_{0}\otimes I,
\end{eqnarray} 
for a one-dimensional position $Q_{0}$ and momentum $P_{0}$.
Then, the input-output relation can be, up to local unitary
equivalence, represented as
\begin{eqnarray} 
M(\De t)&=&Q(0)+(I\otimes Q_{0}\otimes I),\\ 
P(\De t)&=&P(0)-(I\otimes P_{0}\otimes I).
\end{eqnarray} 
Thus, every measurement with independent intervention
has the position error and momentum disturbance 
equivalent with the von Neumann model.

\section{Violations of Heisenberg's relation}
\label{se:VHR}

For finitely accessible input states,  i.e., $\si(Q), \si(P)<\infty$,
\Eq{Ozawa} excludes the possibility of having both $\ep(Q)=0$ 
and $\et(P)=0$ simultaneously.
However, it is possible to have $\et(P)=0$ uniformly over every
input state or alternatively to have $\ep(Q)=0$ uniformly. 
In both cases, \Eq{Heisenberg} is violated uniformly with
$\ep(Q)\et(P)=0$.  
Thus, we have two types of uniform violations of \Eq{Heisenberg};
we shall refer to the former as {\em type I} and the latter as {\em type II}. 

In type I violations, by substituting $\et(P)=0$ in \Eq{Ozawa}, we have
\begin{equation}\label{eq:typeI}
\ep(Q)\si(P)\ge\frac{\hbar}{2}.
\end{equation}
The above relation even allows to have $\ep(Q)\to 0$ with $\si(P)\to\infty$,
and actually a model in Section \ref{se:TypeI} will realize relations $\et(P)=0$ and
$\ep(Q)\to 0$ with $\si(P)\to\infty$.
In this case, the small error is compensated 
not by the large momentum
disturbance but by the large initial momentum uncertainty.
From \Eq{Kennard}, this means that without disturbing the 
momentum the position can be measured as precisely as
our initial knowledge on the object position.  This rather natural
possibility has been excluded from \Eq{Heisenberg}.

Similarly, in type II violations, $\si(Q)$ and $\et(P)$ are 
constrained as 
\begin{equation}\label{eq:typeII}
\si(Q)\et(P)\ge\frac{\hbar}{2},
\end{equation}
so that the small momentum disturbance is compensated 
by the large initial position uncertainty,
and actually a model in Ref.~\cite{02KB5E}
realizes relations $\ep(Q)=0$ and $\et(P)\to 0$ with $\si(Q)\to \infty$.
From \Eq{Kennard}, this implies the possibility of 
the precise position measurement with only disturbing the
momentum as much as the initial momentum uncertainty.
Since \Eq{Heisenberg} has prohibited the precise position 
measurement without infinite momentum transfer, 
this opens a new possibility of precision measurements of
the mass position and similar physical quantities. 

\section{Type I violations and EPR thought experiments}
\label{se:TypeI}

A Type I violation can be obtained by reformulating the 
Einstein-Podolsky-Rosen thought experiment \cite{EPR35}.
Let the measured system be a two-particle system comprising
one-dimensional particles 1 and 2 with positions $Q_{1},Q_{2}$ and
momenta
$P_{1},P_{2}$, respectively, and consider the following process of
measuring $Q_{1}$; our measuring apparatus is assumed to couple only to 
particle 2 and to precisely measure $Q_{2}$, but then to output 
this measured value of $Q_{2}$ as the outcome of the indirect
$Q_{1}$ measurement.  
This is generally not a good measurement of $Q_{1}$; 
however, the interaction for this  measurement does not disturb
$P_{1}$,  so that $\et(P_{1})=0$ uniformly.  
In this case, we can show that
$\ep(Q_{1})^{2}=\bracket{(Q_{1}-Q_{2})^{2}}$  for any input state
\cite{note-3}.
On the other hand, for any small $\al>0$, we can choose a two-particle
state $\ps$ such that $\bracket{(Q_{1}-Q_{2})^{2}}<\al^{2}$. 
Thus, in all such states, we can measure $Q_{1}$ with $\ep(Q_{1})<\al$ 
without disturbing $P_{1}$.
 
The importance of the above example is the
abundance of such state $\ps$.
Let ${\mathcal H}_{i}$ be the state space
of particle $i$ for $i=1,2$.
There exist a unitary operator $U$ on the space 
${\mathcal H}_{1}\otimes{\mathcal H}_{2}$
and a state $\et'\in{\mathcal H}_{2}$ such that for every
$\et$ the state $\ps=U(\et\otimes\et')$ satisfies the condition
$\bracket{(Q_{1}-Q_{2})^{2}}<\al^{2}$.

\section{Type II violations}

All the type II violations are characterized by the condition
$N(Q)=0$. 
Thus, the input-output relation of such measurement is 
characterized by
\begin{eqnarray}
M(\De t)&=&Q(0),\label{eq:Type-II}\\
P(\De t)&=&P(0)+D(P).
\end{eqnarray}

It is generally accepted that every measurement is associated 
with a probability operator-valued measure (POVM) $\Pi$, which 
maps every interval $\De$ to a positive operator $\Pi(\De)$ 
on the state space of the measured object
\cite{Dav76,84QC}; the POVM determines the
probability of obtaining the outcome $\bx$ of the measurement in
an interval $\De$ on input state $\ps$
by $\Pr\{\bx\in\De\}=\bracket{\ps,\Pi(\De)\ps}$.
Then, by \Eq{Type-II} the measurement is 
of type II violation if and only if for any interval
$\De$ the operator $\Pi(\De)$ is the spectral projection of $Q$
corresponding to $\De$.  All the possible state changes associated 
with those measurements of type II violation
were described in Ref.~\cite{01QI}

A model measurement of type II violation arose from the controversy 
on the sensitivity limit of gravitational wave detectors (see \cite{Mad88}
for a brief survey).
The sensitivity limit of interferometer type gravitational wave detectors is
derived from the accuracy of the monitoring of a free mass position, the
relative position of mirrors in the interferometer. 
Braginsky, Caves, and others  \cite{Bra74,CTDSZ80,BK92}
claimed that any position measurement results a back-action
obeying \Eq{Heisenberg} that leads to the following sensitivity 
limit, called the standard quantum limit (SQL),  for monitoring 
the free-mass  position:  {\em Let a free mass $m$ undergo unitary
evolution during the time $\tau $ between two measurements of 
its position $Q$, made with identical measuring apparatuses; the
result of the second measurement cannot be predicted with
root-mean-square error smaller than $(\hbar \tau /m)^{1/2}$ in
average \cite{Cav85}.} 
However, Yuen \cite{Yue83} claimed that the SQL
is not universally valid and can be breached by a
``contractive state measurement'' that leaves the free-mass in a state 
with decreasing position uncertainty in time.
After the debate on the realizability of contractive state 
measurements, it was shown in Ref.~\cite{88MS} 
that the contractive state measurement 
can be realized by the interaction  
 \begin{equation}\label{eq:Ozawa-model}
 H = \frac{K\pi }{3\sqrt3}
\{2(Q\otimes P_{0}-P\otimes Q_{0})
+(QP\otimes I- I\otimes Q_{0}P_{0})\},
\end{equation}
where $Q_{0}$ and $P_{0}$ are the probe position and probe
momentum, respectively.
We can show that this measurement not only breaks the SQL
\cite{88MS} but also is a type II violation of Heisenberg's relation.
In fact, with the meter observable $M=Q_{0}$,  
the time evolution $U=\exp(-i\De tH/\hbar)$ with $K\De t=1$ 
determines the input-output relation \cite{02KB5E}
\begin{eqnarray}
M(\De t)&=&Q(0),\\
P(\De t)&=&P(0)+[P_{0}(0)-P(0)].
\end{eqnarray}
From the above input-output relation, 
we have $N(Q)=0$ and $D(P)=P_{0}(0)-P(0)$,
so that $\ep(Q)=0$ uniformly and
$\et(P)^{2}=\bracket{P_{0}(0)^{2}}+\bracket{P(0)^{2}}
-2\bracket{P_{0}(0)}\bracket{P(0)}<\infty$.
Thus, $\ep(Q)\et(P)=0$ uniformly and the above measurement
is of type II violation of \Eq{Heisenberg}.
For more detail on this measurement, we refer to 
Refs.~\cite{88MS,89RS,01CQSR,02KB5E,03UVRHUP,90QP}.

\section{Concluding remarks}

Many text books have associated the formal expression of  
``Heisenberg's uncertainty relation'' to \Eq{Kennard}, 
but also associated the physical meaning of 
``Heisenberg's uncertainty relation'' to \Eq{Heisenberg}.   
However, \Eq{Kennard} can no longer be considered 
as a formal expression of \Eq{Heisenberg}, 
since \Eq{Kennard} is a universally valid mathematical relation,   
but \Eq{Heisenberg} has its own limitation.

In this Letter, we define ``Heisenberg's uncertainty relation'' 
based on Heisenberg's claim published in 1927, 
which, of course, revealed the limitation of our ability 
of observation on microscopic objects at the first time.  
However, Heisenberg himself appears to have changed his position 
from 1927 to 1929. 
Around this time, it was already known that the EPR type thought
experiment violates \Eq{Heisenberg}. 
In this case, the position of the mass at a time $t$ can be indirectly
measured very precisely without disturbing the momentum, 
and hence if the momentum is measured directly just after the
position measurement, the momentum at the time $t$ can also be 
measured very precisely, as discussed in Section \ref{se:TypeI}. 
Heisenberg's  response to this criticism appears to have been
that he considered the uncertainty relation to be \Eq{Kennard} 
rather than \Eq{Heisenberg}  by stating, for instance, that ``every
experiment destroys some of the  knowledge of  the system which was
obtained by previous experiments.     
This formulation makes it clear that the uncertainty relation does not 
refer to  the past''  (p. 20, Ref.~\cite{Hei30}).    
Heisenberg's response means that even if we can measure 
both the position and the momentum at the past time $t$ very precisely, 
after the momentum measurement the mass has no longer 
definite position so that \Eq{Kennard} is not violated at any time. 
Thus, around this time, ``Heisenberg's uncertainty relation'' might 
turn to be a more fundamental formal relation like the CCR than what 
Heisenberg claimed in 1927.

Although there have been many attempts to show violations of 
\Eq{Heisenberg}, 
it has been difficult to construct a forceful argument without definite 
definitions of error and disturbance, since, as shown in 
Section \ref{se:VHR},
any such conceivable violation has an alternative compensating 
reciprocal relation,
\Eq{typeI} or \Eq{typeII}, with the same lower bound.
For instance, if one shows a calculation on some model indicating 
that \Eq{Heisenberg}
is ultimately violated, then the defender can easily modify 
the calculation to lead
to a reciprocal relation \Eq{typeI} or \Eq{typeII} 
and tell that not only \Eq{Kennard}
but also ``Heisenberg's uncertainty relation''  holds in this measurement.

Thus, the definiteness of the definition of error is quite important 
in the whole story.
We can justify our definition of error as follows.
If the input state has the true value of the measured observable, 
our definition gives the root-mean-square difference 
between the true value and the measured value, 
and hence this is the only one from the name ``root-mean-square error.''
We can also show that under some natural mathematical requirements, 
extending the definition to arbitrary input states is also unique.
In this Letter, the root-mean-square error is defined through 
the model of measuring process. 
However, this error can be shown to be model independent and actually
determined directly from the POVM of the measurement.  
In fact, the error $\ep(A)$ in measuring an observable $A$ using 
apparatus with a POVM $\Pi(dx)$ coincides with the distance $d(\Pi,A)$
between the POVM $\Pi(dx)$ and the observable $A$ defined by
\begin{eqnarray}
d(\Pi,A)^{2}&=&\left\bracket{\ps\left|
\int x^{2}\,\Pi(dx)-\left[\int x\,\Pi(dx)\right]^{2}\right|\ps\right}\nn\\
& &\mbox{}+\left\|\int x\,\Pi(dx)\ps - A\ps\right\|^{2}.
\end{eqnarray}
The detail will be discussed in a forthcoming paper \cite{03NDM}.
According to that, the definition of root-mean-square error is 
believed to be unique. 

The $\gamma$ ray microscope is considered to be described by a
model with independent intervention under usual approximation 
allowed in mathematical modeling of the physical apparatus. 
Then, the
theory shows that the root-mean-square error and the
root-mean-square momentum disturbance satisfy Heisenberg's
relation. The resolution power of the $\gamma$ ray microscope is
considered to  coincide with the experimental root-mean-square
error in the state with definite position before measurement up to
constant factor. Thus, the resolution power should coincide with the
root-mean-square error of the model up to constant factor, so that
the theory consistently concludes that the product of the resolution
power  and the momentum disturbance is lower bounded by 
the Planck constant times a constant factor.

In the present Letter, we have shown  that Heisenberg's reciprocal
relation between position-measurement error and momentum
disturbance holds for every measurement with independent
intervention and clarifies the limitation of Heisenberg's relation. We
also proposed a new universally valid relation among measurement
error,  disturbance, and initial uncertainties.  This relation reveals
possibilities of measurements beyond Heisenberg's relation and
clarifies the new constraints for measurements beyond Heisenberg's relation.  
 
\section*{Acknowledgments} 

This work was supported by the 
Strategic Information and Communications R\&D Promotion Scheme
of the MPHPT of Japan,  by the CREST
project of the JST, and by the Grant-in-Aid for Scientific Research of
the JSPS.


\begin{thebibliography}{99}

\bibitem{Boh28}
N.~Bohr,
{Nature (London)} {\bf 121}, 580 (1928).

\bibitem{Hei27}
W.~Heisenberg, Z. Phys.\ {\bf 43}, 172 (1927)
[in Quantum Theory and Measurement, 
edited by J.~A. Wheeler and
W.~H. Zurek, pages~62--84 (Princeton
University Press, 1983)].

\bibitem{Ken27}
E.~H. Kennard,
{Z. Physik} {\bf 44}, 326 (1927).

\bibitem{Rob29}
H.~P. Robertson,
{Phys.\ Rev.} {\bf 34}, 163 (1929).

\bibitem{note-1}
In this Letter, every state vector is assumed normalized and
the domain of the commutator $[A,B]$ is considered extended
appropriately.

\bibitem{vN32}
J.~von Neumann,
Mathematische Grundlagen der
Quantenmechanik (Springer, Berlin, 1932)
[Mathematical Foundations of Quantum Mechanics
(Princeton University Press, 1955)].

\bibitem{Boh49}
N.~Bohr,
in {Albert Einstein: Philosopher-Scientist}, 
edited by P.~A. Shilpp, pages 200--241
(Evanston, The Library of Living Philosophers, 1949).

\bibitem{Boh51}
D.~Bohm, {Quantum Theory}
(Prentice-Hall, New York, 1951).

\bibitem{Mes59a}
A.~Messiah, M\'{e}canique Quantique, I (Dunod, Paris, 1959),
[{Quantum Mechanics}, I  (North-Holland, Amsterdam, 1959)].

\bibitem{Bra74}
V.~B. Braginsky and Yu.\ I. Vorontsov,
{Uspehi Fiz.\ Nauk} {\bf 114}, 41 (1974)
[Sov.\ Phys.\ Usp.\ {\bf 17}, 644 (1975)].

\bibitem{CTDSZ80}
C.~M. Caves {\em et al.}, {Rev.\ Mod.\ Phys.} {\bf 52}, 341 (1980).

\bibitem{BVT80}
V. B. Braginsky {\em et al.}, Science {\bf 209}, 547 (1980).

\bibitem{EPR35}
A.~Einstein, B.~Podolsky, and N.~Rosen,
{Phys.\ Rev.} {\bf 47}, 777 (1935).

\bibitem{AK65}
E.~Arthurs and J.~L. {Kelly, Jr.},
{Bell.\ Syst.\ Tech.\ J.} {\bf 44}, 725 (1965).

\bibitem{Bal70}
L.~E. Ballentine,
{Rev.\ Mod.\ Phys.} {\bf 42}, 358 (1970).

\bibitem{Yue83}
H.~P. Yuen,
{Phys.\ Rev.\ Lett.} {\bf 51}, 719 (1983).

\bibitem{Kra87}
K.~Kraus.
{Phys.\ Rev.\ D} {\bf 35}, 3070 (1987).

\bibitem{AG88}
E.~Arthurs and M.~S. Goodman,
{Phys.\ Rev.\ Lett.} {\bf 60}, 2447 (1988).

\bibitem{88MS}
M.~Ozawa,
{Phys.\ Rev.\ Lett.} {\bf 60}, 385 (1988).

\bibitem{89RS}
M.~Ozawa,
in {Squeezed and Nonclassical
  Light}, edited by P.~Tombesi and E.~R. Pike, 
pages 263--286 (Plenum, New York, 1989).

\bibitem{HV90}
J.~Hilgevoord and J.~Uffink,
in {Sixty-Two Years of Uncertainty},  edited by A.~I. Miller,
pages 121--137 (Plenum, New York, 1990). 

\bibitem{MM90}
H.~Martens and W.~M. de~Muynck,
{Found.\ Phys.} {\bf 20}, 357 (1990).

\bibitem{91QU}
M.~Ozawa,
in {Quantum Aspects of Optical Communications}, 
edited by C.~Bendjaballah {\em et al.}, pages 3--17
(Springer, Berlin, 1991).

\bibitem{Ish91}
S.~Ishikawa,
{Rep.\ Math.\ Phys.} {\bf 29}, 257 (1991).

\bibitem{BK92}
V.~B. Braginsky and F.~Ya.\ Khalili,
{Quantum Measurement},
(Cambridge University Press, Cambridge, 1992).

\bibitem{MM92}
H.~Martens and W.~M. de~Muynck,
{J. Phys.\ A} {\bf 25}, 4887 (1992).

\bibitem{App98}
D.~M. Appleby, Int.\ J. Theor.\ Phys.\ {\bf 37}, 1491 (1998).

\bibitem{01CQSR}
M.~Ozawa,
{Phys.\ Lett.\ A} {\bf 282}, 336 (2001).

\bibitem{02KB5E}
M.~Ozawa,
{Phys.\ Lett.\ A} {\bf 299}, 1 (2002).

\bibitem{03UVRHUP}
M.~Ozawa,
{Phys.\ Rev.\ A} {\bf 67}, 042105 (2003).


\bibitem{note-2}
In the above proof, we illegitimately assume that 
the input state is a momentum eigenstate, and that 
the output state is a well-defined pure state.  
The argument can be improved so that
the input state is assumed to be a normalized state 
arbitrarily close to the momentum eigenstate.
Then, the output state is generally well-defined as a density operator.  
Mathematically complete argument along with this line is more
involved and will be given elsewhere.

\bibitem{RS75}
M.~Reed and B.~Simon,
{Methods of Modern Mathematical Physics, II: {Fourier} Analysis,
  Self-Adjointness}.
(Academic, New York, 1975).

\bibitem{note-3}
Since $Q_{2}$ is measured precisely, we have 
$N(Q_{2})=M(\De t)-Q_{2}(0)=0$, so that 
$N(Q_{1})=M(\De t)-Q_{1}(0)=Q_{2}(0)-Q_{1}(0)$, and hence
we have $\ep(Q_{1})^{2}=\bracket{(Q_{1}-Q_{2})^{2}}$.

\bibitem{Dav76}
E. B. Davies, Quantum Theory of Open Systems,
(Academic, London, 1976).

\bibitem{84QC}
M. Ozawa, J. Math.\ Phys. {\bf 25}, 79 (1984).

\bibitem{01QI}
M. Ozawa, in Quantum Communication, Computing, and 
Measurement 3, edited by P. Tombesi and O. Hirota,
pages  97--106 (Kluwer/Plenum, New York, 2001).

\bibitem{Mad88}
J. Maddox, Nature (London) {\bf 331}, 559 (1988).

\bibitem{Cav85}
C.~M. Caves, Phys.\ Rev.\ Lett. {\bf 54}, 2465 (1985).

\bibitem{90QP}
M. Ozawa, Phys.\ Rev.\ A {\bf 41}, 1735 (1990).

\bibitem{Hei30}
W. Heisenberg, The Physical Principles of the Quantum Theory,
(University of Chicago Press, Chicago, 1930)
[Reprinted by Dover, New York (1949, 1967)]. 

\bibitem{03NDM}
M. Ozawa, in preparation.

\end{thebibliography}
\end{document}